# Deprojection of Axially Symmetric Objects

Christopher S. Kochanek

and

George B. Rybicki

Harvard-Smithsonian Center for Astrophysics

Received ____________________; accepted ____________________

## ABSTRACT

The deprojection of axisymmetric density distributions is generally indeterminate to within the addition of certain axisymmetric distributions (konus densities) that are invisible in projection. The known class of konus densities is expanded considerably here through the introduction of *semikonus functions*. These functions are closed with respect to multiplication in ordinary space, and the real part of an arbitrary polynomial of semikonus functions is a konus function. This property facilitates the construction of semikonus (and konus) functions with tailored properties, such as asymptotic forms. We also develop a simple technique for constructing several classes of konus distributions with arbitrary density profiles in the equatorial plane.

## 1. Introduction

The deprojection of objects, such as galaxies, clusters or X-ray emission regions, is a common problem in astronomy. For the case of spherically symmetric objects, formal deprojection by means of Abel transforms is well known (Courant & Hilbert 1962). The general problem of deprojection of axisymmetric objects was considered by Rybicki (1986), who showed, by use of the Fourier slice theorem, that deprojection is indeterminate in principle unless the angle of inclination $i = 90°$. In particular, the deprojection is indeterminate by the addition of any function that vanishes outside a "cone of ignorance" in Fourier space, with its axis aligned with the symmetry axis of the object and with half-angle $90° - i$. Such distributions have been called "konus densities" by Gerhard & Binney (1995; hereafter GB).

It is important to study the range of possible konus distributions to understand the uncertainties in the structure of observed axisymmetric objects. In their paper GB investigated a single konus distribution constructed from exponentials in cylindrical coordinates in Fourier space. In this paper we develop several methods of constructing konus distributions based on two ideas.

The first idea is the related concept of a *semikonus* function. These are similar to konus functions, except they vanish in Fourier space outside a half-cone rather than a full cone. Given a semikonus function, it is easy to construct a konus function by adding an appropriate reflected function in Fourier space. This is equivalent to taking the real part of the function in ordinary space. The advantage of semikonus functions is that they can be simply combined to form new semikonus functions. Specifically, the class of semikonus functions is shown to be closed under multiplication in ordinary space (convolution in Fourier space). For example, by simply raising one known semikonus function to arbitrary powers, one can construct a whole set of semikonus functions. This allows the construction of semikonus (and konus) functions that decrease arbitrarily rapidly at large radii.

The second idea is a simple method for generating semikonus functions with arbitrary profiles in the equatorial plane of the distribution. We define a kernel for generating semikonus functions by the Fourier transform of a single spherical function vanishing outside a sphere, and then translate it up a distance $k$ along the $k_z$ axis until it lies inside a cone. Weighted averages of such kernels for a range of values of $k$ produce semikonus functions in terms of sine and cosine transforms of the weighting function.

We develop the theory of semikonus functions in §2, and derive particular classes of solutions in §3. In §4 we discuss the implications of these results for the interpretation of observed axisymmetric objects.

## 2. Konus and Semikonus Functions

We adopt the notation of GB. The space density $\rho(\mathbf{r})$ of the object is related to its Fourier transform $A(\mathbf{k})$ by

$$\rho(\mathbf{r}) = \frac{1}{(2\pi)^3} \int d^3\mathbf{k}\, A(\mathbf{k}) \exp(i\mathbf{k}\cdot\mathbf{r}). \tag{1}$$

The inverse of this is

$$A(\mathbf{k}) = \int d^3\mathbf{r}\, \rho(\mathbf{r}) \exp(-i\mathbf{k}\cdot\mathbf{r}). \tag{2}$$

By assumption the object is axisymmetric. Without loss of generality, the $z$-axis can be taken as the axis of symmetry, with unit vector $\hat{\mathbf{z}}$. In Fourier space $\mathbf{k}$, consider the cone $\mathcal{C}(\theta)$ with axis in direction $\hat{\mathbf{z}}$ with half-angle equal to $\theta$. A function whose Fourier transform $A(\mathbf{k})$ vanishes outside this cone has the property that the corresponding space density $\rho(\mathbf{r})$ is invisible in projection for any inclination angle $i$ less than $90° - \theta$. Rybicki (1986) calls this the "cone of ignorance", and GB call any such function a "konus" density.

It is very useful to sharpen the concept of a konus density. The cone $\mathcal{C}(\theta)$ consists of two half-cones, denoted by $\mathcal{C}_+(\theta)$ and $\mathcal{C}_-(\theta)$, that extend in the positive and negative $\hat{\mathbf{z}}$ directions, respectively. We call a distribution function whose Fourier transform vanishes outside one of these half-cones a *semikonus* function. The distinction between the semikonus functions defined with respect to the positive and negative half-cones is given by an appropriate subscript, as $A_+(\mathbf{k})$ or $A_-(\mathbf{k})$.

The relationship between the above semikonus functions and the konus distributions defined by GB is simple. Any konus distribution can be represented as

$$A(\mathbf{k}) = \frac{1}{2}\left[A_+(\mathbf{k}) + A_-(\mathbf{k})\right]. \tag{3}$$

In order that $\rho(\mathbf{r})$ be real, the following relation must hold,

$$A_-(\mathbf{k}) = A_+^*(-\mathbf{k}), \tag{4}$$

where $^*$ denotes complex conjugation. The corresponding result for the spatial konus function is

$$\rho(\mathbf{r}) = \Re\rho_+(\mathbf{r}), \tag{5}$$

where $\Re f$ denotes the real part of $f$.

The importance of the concept of semikonus functions is that they are closed under convolution in Fourier space. That is, if $A_+^{(1)}(\mathbf{k})$ and $A_+^{(2)}(\mathbf{k})$ are semikonus functions with respect to the half-cone $\mathcal{C}_+(\theta)$, so also is their convolution

$$A_+(\mathbf{k}) = A_+^{(1)}(\mathbf{k}) * A_+^{(2)}(\mathbf{k}) = \int d^3\mathbf{k}'\, A_+^{(1)}(\mathbf{k}-\mathbf{k}')A_+^{(2)}(\mathbf{k}'). \tag{6}$$

To prove this, first note that the condition that a vector $\mathbf{k}$ lies within $\mathcal{C}_+(\theta)$ can be written $\mathbf{k}\cdot\hat{\mathbf{z}} > \mu|\mathbf{k}|$, where $\mu = \cos\theta$. Now, if two vectors $\mathbf{k}_1$ and $\mathbf{k}_2$ each lie within the cone $\mathcal{C}_+(\theta)$, so does their sum $\mathbf{k} = \mathbf{k}_1 + \mathbf{k}_2$, since

$$\mathbf{k}\cdot\hat{\mathbf{z}} = \mathbf{k}_1\cdot\hat{\mathbf{z}} + \mathbf{k}_2\cdot\hat{\mathbf{z}} > \mu\left(|\mathbf{k}_1| + |\mathbf{k}_2|\right) > \mu|\mathbf{k}|, \tag{7}$$

where the triangle inequality has been used. Applying this result to Eq. (6), if $\mathbf{k}$ lies outside the cone, then at least one of the two vectors $\mathbf{k}-\mathbf{k}'$ or $\mathbf{k}'$ must lie outside the cone, implying that at least one of the two functions $A_+^{(1)}(\mathbf{k}-\mathbf{k}')$ or $A_+^{(2)}(\mathbf{k}')$ vanishes for any value of $\mathbf{k}'$. Thus the integral vanishes, establishing that $A_+(\mathbf{k})$ is a semikonus function. Analogous results apply to semikonus functions with respect to the negative cone.

The preceding result can be trivially generalized to cases where the two functions $A_+^{(1)}$ and $A_+^{(2)}$ may have different cone angles, $\theta^{(1)}$ and $\theta^{(2)}$. Since a semikonus function with a given value of $\theta$ is a semikonus function for any larger value of $\theta$ as well, it follows that their convolution is a semikonus function with $\theta = \max(\theta^{(1)}, \theta^{(2)})$.

It follows from the Fourier convolution theorem that if both $\rho_+^{(1)}(\mathbf{r})$ and $\rho_+^{(2)}(\mathbf{r})$ are the spatial representations of semikonus functions, then their ordinary product,

$$\rho_+(\mathbf{r}) = \rho_+^{(1)}(\mathbf{r}) \cdot \rho_+^{(2)}(\mathbf{r}), \tag{8}$$

is a semikonus function with respect to the same half-cone. That is, semikonus spatial functions are closed under ordinary multiplication. This provides a simple way to construct new semikonus functions from other semikonus functions. In particular, applying the result $n$ times to the same function $\rho_+^{(1)}(\mathbf{r})$ gives the new semikonus function

$$\rho_+^{(n)}(\mathbf{r}) = \left[\rho_+^{(1)}(\mathbf{r})\right]^n . \tag{9}$$

The konus distributions corresponding to Eqs. (8) and (9) are

$$\rho(\mathbf{r}) = \Re\left[\rho_+^{(1)}(\mathbf{r}) \cdot \rho_+^{(2)}(\mathbf{r})\right], \tag{10}$$

and

$$\rho^{(n)}(\mathbf{r}) = \Re\left[\rho_+^{(1)}(\mathbf{r})\right]^n . \tag{11}$$

Since semikonus spatial functions are complex, multiplication and taking the real part are noncommuting operations, so it is *not* true, for example, that $\rho^{(n)}(\mathbf{r}) = [\rho^{(1)}(\mathbf{r})]^n$.

The imaginary parts of semikonus functions are essential to the closure property under multiplication.

Eq. (9) provides a simple method for constructing semikonus functions (and konus functions) that decay rapidly with $r = |\mathbf{r}|$. For example, if $\rho_+^{(1)}(\mathbf{r}) = O(1/r^\beta)$, then $\rho_+^{(n)}(\mathbf{r}) = O(1/r^{n\beta})$. The complex phase of $\rho_+^{(n)}$ is increased by a factor $n$ in this process, so the associated konus density $\rho^{(n)}$ will show a greater number of sign alterations in addition to the more rapid asymptotic decay.

## 3. General Families of Konus and Semikonus functions

We now shall construct several families of analytic semikonus and konus functions. Although they are quite simple, they provide useful examples of the kinds of indeterminacies that limit the deprojection of axisymmetric objects. We shall use the spherical coordinate $r$ as well as the cylindrical coordinates $z$ and $R = (r^2 - z^2)^{1/2}$, sometimes in various combinations.

Let $f(k, k_0)$ be a spherical function in Fourier space that vanishes for all $k > k_0$, with corresponding real space density

$$F(r, k_0) = \frac{1}{2\pi^2} \int_0^{k_0} k^2 dk \frac{\sin kr}{kr} f(k, k_0). \tag{12}$$

We call the Fourier transform pair $f(k, k_0) \leftrightarrow F(r, k_0)$ a kernel for producing a family of semikonus functions. If we translate a kernel along the $k_z$ axis by distance $|k| = k_0/\sin\theta$, then the non-zero region of the kernel lies entirely inside a cone of half-angle $\theta$, and the density function

$$\rho_+(r, z|k, \theta) = F(r, k\sin\theta)e^{-ikz} \tag{13}$$

is a semikonus function that is invisible at inclinations $i < 90° - \theta$. In general $\rho_+(r, z|k, \theta)$ is rapidly oscillating, but we can smooth it using the fact that any weighted sum of semikonus functions is also a semikonus function. If $g(k)$ is a weighting function, then we can build the smoothed semikonus function

$$\rho_+(r,z|\theta) = \int_0^\infty dk g(k) e^{-ikz} F(r, k\sin\theta). \qquad (14)$$

By construction, this is a semikonus function for any choice of $f(k,k_0)$ and $g(k)$.

To facilitate evaluating the integrals in Eq. (14) we would like to have a simple form for the kernel function $F(r,k_0)$. The simplest kernel is a shell of radius $k_0$ in Fourier space, $f(k,k_0) = \delta(k-k_0)/k_0$. The real space kernel is $F(r,k_0) = \sin(k_0 r)/2\pi^2 r$, and Eq. (14) simplifies to

$$\rho_+(r,z|\theta) = \frac{1}{2\pi^2 r}\int_0^\infty dk g(k) e^{-ikz}\sin(kr\sin\theta). \qquad (15)$$

Using appropriate trigonometric identities, the integral can be expressed as a sum of Fourier sine and cosine transformations of the weighting function $g(k)$. If we define

$$\alpha_\pm = r\sin\theta \pm z, \qquad (16)$$

and the sine and cosine transforms

$$f_c(x) = (4\pi^2)^{-1}\int_0^\infty dk g(k)\cos(kx) \quad \text{and} \quad f_s(x) = (4\pi^2)^{-1}\int_0^\infty dk g(k)\sin(kx), \qquad (17)$$

then

$$\rho_+(r,z|\theta) = \frac{[f_s(\alpha_+) + f_s(\alpha_-)]}{r} - i\frac{[f_c(\alpha_+) - f_c(\alpha_+)]}{r}. \qquad (18)$$

Given any odd function $f_s(-x) = -f_s(x)$, the function $\rho(r,z|\theta) = \Re\rho_+(r,z|\theta)$ is a konus function. The density profile of this solution in the equatorial plane ($z=0$) is $P(R) = \rho(R,0|\theta) = 2f_s(R\sin\theta)/R$. In terms of this equatorial profile, the konus distribution throughout all space can be written

$$\rho(r,z|\theta) = \frac{1}{r}\left[\hat\alpha_+ P\left(\hat\alpha_+\right) + \hat\alpha_- P\left(\hat\alpha_-\right)\right] \qquad (19)$$

where $\hat\alpha_\pm = \alpha_\pm/\sin\theta$. Unfortunately, there is no simple formula for the imaginary part of the semikonus density, $\Im\rho_+$, in terms of the equatorial profile used to define the real part.

The function $\Im\rho_+(r,z|\theta)$ is also a konus function for any even function $f_c(-x) = f_c(x)$, but it is antisymmetric in $z$ and therefore unsuitable by itself for representing typical galaxies (although it may be useful in other contexts). Knowledge of the imaginary part is nonetheless important here, because the closure under multiplication only holds for complete semikonus functions.

Although the projection properties of the solution in Eq. (18) follow from the general theory presented above, it is of some interest to sketch a direct proof from the projection integral. This can be done by changing the variables in the projection integral to $d\alpha_+$ and $d\alpha_-$ for the functions of $\alpha_+$ and $\alpha_-$ respectively. For viewing angles $i > 90° - \theta$ the integration limits are asymmetric and extend from a finite value to $\pm\infty$ and cancellation cannot occur. For viewing angles $i < 90° - \theta$ the integration limits extend from $-\infty$ to $+\infty$, and the contributions of the first and second terms in $\Re\rho_+$ (or $\Im\rho_+$) cancel.

We can immediately use $\Re\rho_+(r,z|\theta)$ as an example of a konus function for any odd function $f_s(x)$. If the defining function $f_s(x)$ asymptotically declines as $r^{-n}$, then $\Re\rho_+$ will asymptotically decline as $r^{-n-1}$ except near the curves $\alpha_\pm = 0$. Along a curve with $\alpha_\pm$ held constant, the function only declines as $r^{-1}$. If $z = r\sin\phi$ then holding $\alpha_\pm = c$ fixed for $r \gg 1$ implies that $\sin\phi = \mp\sin\theta \pm c/r$, so the region where the function declines as $r^{-1}$ is confined to an asymptotically vanishing region near the $\alpha_\pm = 0$ curves. A function $f_s(\alpha_\pm)$ with no extrema at finite values of $\alpha_\pm$ has no asymptotic anomalies, but such functions cannot be used to model galaxies because they must have divergent densities as $|\alpha_\pm| \to \infty$. The imaginary part of the solution shows the same problem. Faster asymptotic decline can be obtained by taking powers of this solution, but asymptotic anomalies are still prominent along these special curves, at least relatively.

The $n$th power of the semikonus function in Eq. (18), $\rho_+^n(r,z|\theta)$, is also a semikonus function. This function declines as $r^{-n}$ near the curves $\alpha_\pm = 0$, and as $r^{-nm}$ away from the curves if $\rho_+$ asymptotically declines as $r^{-m}$. Alternatively, if we chose the same functions but use solutions with two different cones of uncertainty $\theta$ and $\theta'$ ($\theta \neq \theta'$), then $\rho_+(r,z|\theta)\rho_+(r,z|\theta')$ asymptotically decreases as $r^{-1-n}$ near the two degenerate curves, and as $r^{-2n}$ away from the degenerate curves. The somewhat pathological asymptotic behavior along the $\alpha_\pm = 0$ curves appears to be characteristic of the konus functions due to non-analyticity along the cone in Fourier space (the GB solution also shows this asymptotic property).

A particularly simple result is found for the exponential weight function $g(k) \propto \exp(-bk)$. In this model $f_c = b/(b^2+x^2)$ and $f_s = x/(b^2+x^2)$ so that

$$\rho_{1+}(r,z|\theta) = \frac{2\sin\theta}{r^2\sin^2\theta - (z+ib)^2}, \tag{20}$$

and

$$\Re\rho_{1+} = 2\sin\theta\frac{b^2+\alpha_+\alpha_-}{(b^2+\alpha_+^2)(b^2+\alpha_-^2)} \quad \text{and} \quad \Im\rho_{1+} = 2i\sin\theta\frac{b(\alpha_+-\alpha_-)}{(b^2+\alpha_+^2)(b^2+\alpha_-^2)} \tag{21}$$

The function $\Re\rho_{1+}$ declines as $r^{-2}$ at all angles away from $\alpha_\pm = 0$, while the function $\Im\rho_{E+}$ declines as $r^{-3}$ away from the lines where $\alpha_\pm = 0$. Figure 1 shows contours of the first four konus functions in the sequence $\rho_n(r,z|\theta) = \Re\rho_{1+}^n(r,z|\theta)$, $n = 1, 2, 3, 4$. Near $\alpha_\pm = 0$ the solutions decline as $r^{-n}$ and elsewhere as $r^{-2n}$, and in the plane the equatorial profile is $\propto (R^2\sin^2\theta + b^2)^{-2n}$. For comparison, Figure 2 shows the solution to Eq. (18) with the same equatorial profile as $\rho_2$. This solution has only a single sign alteration instead of two. Figure 3 shows the angular profile of the function $\rho_1$ at two different radii with $r \gg b$ to illustrate the differences in the rate of convergence. Figure 4 uses the solution $\rho_E(r,z|\theta)$ to Eq. (18) with an exponential density profile in the equatorial plane [$g(k) = 1/(1+b^2k^2)$]. The first two solutions, $\rho_{E1}(r,z|30^\circ)$ and $\rho_{E2}(r,z|30^\circ) = \rho_{E1}^2(r,z|30^\circ)$, converge to zero exponentially except near $\alpha_\pm = 0$, where the convergence reverts to $r^{-1}$ or $r^{-2}$. The last solution, $\rho_{E1}(r,z|30^\circ)\rho_{E1}(r,z|50^\circ)$ which is a konus solution for $\theta = 50^\circ$ by the theorems of §2, converges exponentially at all angles.

Other kernels also lead to simple solutions. For example, if we define the kernel by a filled sphere instead of a shell, then we find the solutions

$$\rho_+(r,z|\theta) = -\frac{1}{r}\left\{\left[\frac{f_s(\alpha_+)}{r}\right]' + \left[\frac{f_s(\alpha_-)}{r}\right]'\right\} - \frac{i}{r}\left\{\left[\frac{f_c(\alpha_+)}{r}\right]' - \left[\frac{f_c(\alpha_-)}{r}\right]'\right\} \tag{22}$$

where the $'$ denotes a derivative with respect to $r$ at fixed $z$ (i.e. $\alpha'_+ = \sin\theta$). If $\Re\rho_+$ is to have the equatorial profile $P(R\sin\theta)$, then $f(x) = (x/2)\int_x^\infty uP(u)du$. This class of solutions declines as $r^{-2}$ along lines of constant $\alpha_\pm$ instead of $r^{-1}$. If we use the weight function $g(k) = \exp(-bk)$ then the solution is identical to the solution $\rho_{1+}^2$ found by using the solution for a spherical shell with the same weight function (Eq. 20) and squaring it. Smoothing the kernel by changing from a spherical shell to a filled sphere is (in this case) equivalent to the convolution of the Fourier transforms of the semikonus functions for the spherical shell. Polynomial kernels for $f(k,k_0)$ produce kernels $F(r,k_0)$ that are polynomials in $k_0$ multiplied by $\sin k_0 r$ or $\cos k_0 r$. These lead to more general forms of Eq. (22) that depend on functions $f(\alpha_\pm)$ and $f(z)$ and their derivatives. We can also find analytic solutions for kernels including a $\sin ak$ (or $\cos ak$) dependence, and in these solutions the $\alpha_\pm$ "coordinates" are modified to $(r \pm a)\sin\theta \pm z$, and the solution will include all four sign combinations. Spatial or parametric derivatives and integrals of semikonus functions are also semikonus functions.

## 4. Conclusions

Far from it being difficult to construct konus distributions, it is easy to find a variety of solutions with arbitrary equatorial profiles. Some of these solutions are surprisingly simple. Given one solution, the closure property under multiplication for semikonus functions allows the construction of sequences of solutions with more rapid asymptotic convergence and increasing numbers of zones with alternating signs. As a result, we can construct functions that decline arbitrarily rapidly at large radii.

Our construction method always produces a different rate of asymptotic decay near the lines $\alpha_\pm = r\sin\theta \pm z = 0$ where $\sin\theta$ is the cone opening angle. Any cut in angle through the density profile at a fixed large radius eventually shows a sharp pair of positive and negative density peaks straddling the curves $\alpha_\pm = 0$. The solution found by GB shows similar asymptotic properties. This pathology is probably caused by the structure we imposed on the integral to find analytic solutions, and it is not obviously a necessary property of konus functions. However, a purely numerical approach appears to be needed to search for konus functions with smoother asymptotic profiles, since we failed to find analytic solutions without these asymptotic properties.

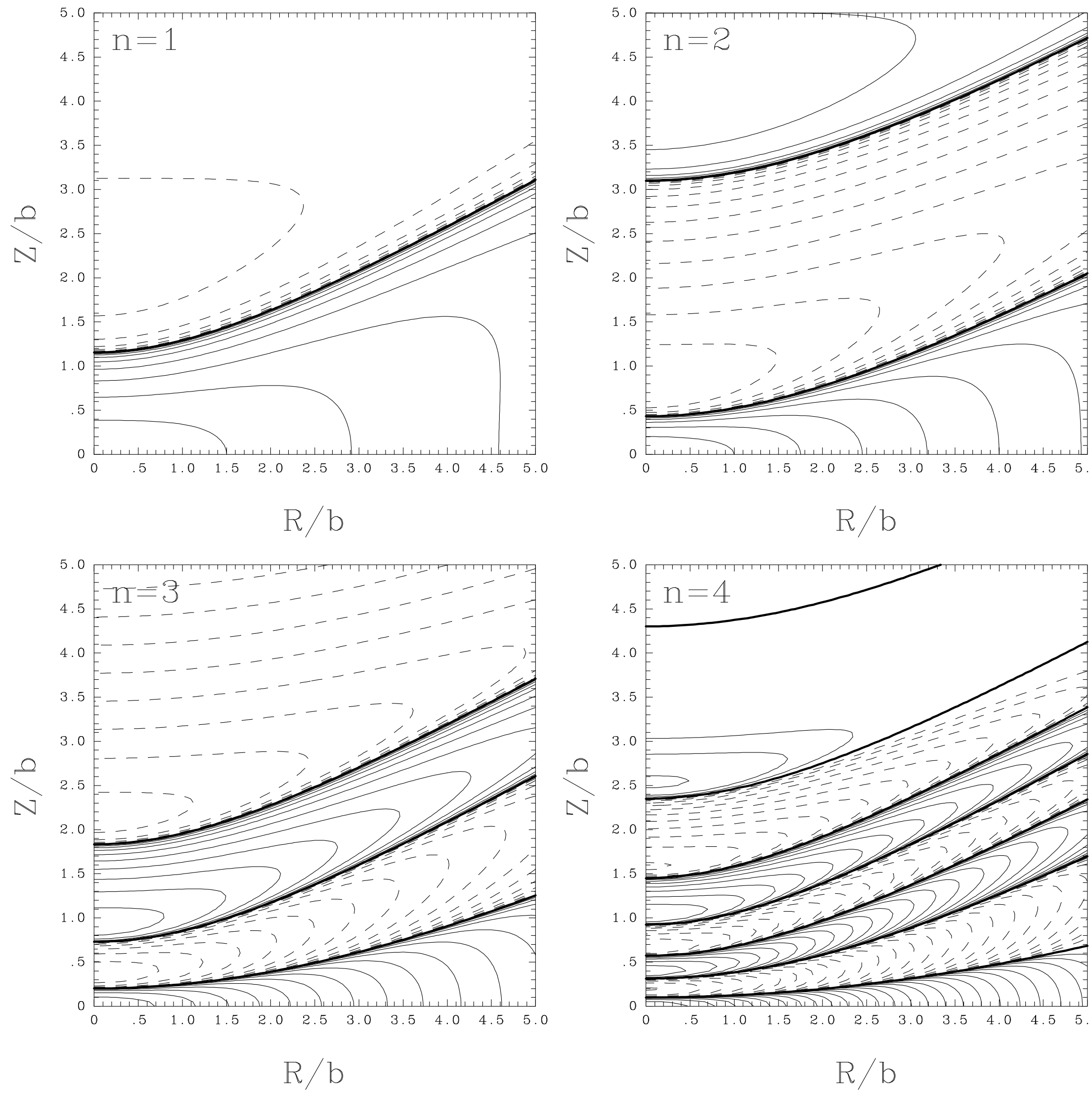

Fig. 1.— Contours of the konus functions $\Re\rho^n_{1+}(r,z|30°)$ for $n=1$ to 4. The opening angle of the cone of uncertainty is $\theta=30°$, and each function is normalized to have a peak value of unity. In projection these functions are invisible for $i<60°$. The contours start at $\pm64\%$ of the peak and there is a factor of two reduction between each contour. Positive contours are solid, negative contours are dashed. Note that the asymptotic decay for large radii becomes faster for larger values of $n$ ($\propto r^{-2n}$), but there are more sign alterations.

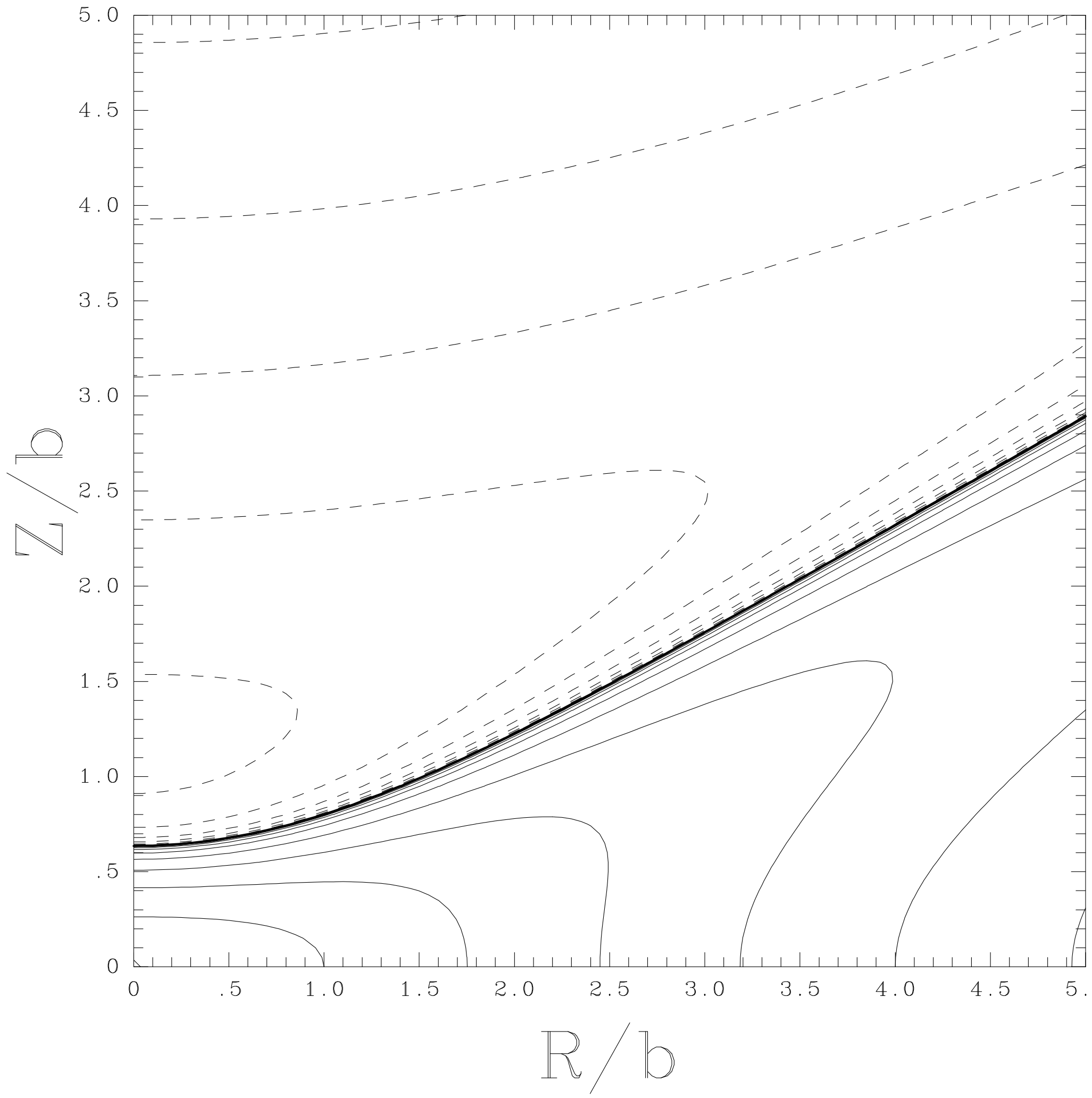

Fig. 3.— The solution to Eq. (15) with the equatorial density profile $(R^2 \sin\theta^2 + b^2)^{-2}$ and $\theta = 30°$. In the equatorial plane this solution has the same density profile as the solution $\Re\rho^2_{1+}$ shown in the top right panel of Figure 1. Away from the curves $\alpha_\pm = 0$ both solutions decay as $r^{-4}$, but near $\alpha_\pm = 0$ this solution decays as $r^{-1}$ while the $\rho^2_{1+}$ solution decays as $r^{-2}$. Note also the single sign change in this solution compared to the two sign changes in the $\rho^2_{1+}$ solution. Scaling and contours are the same as in Figure 1.

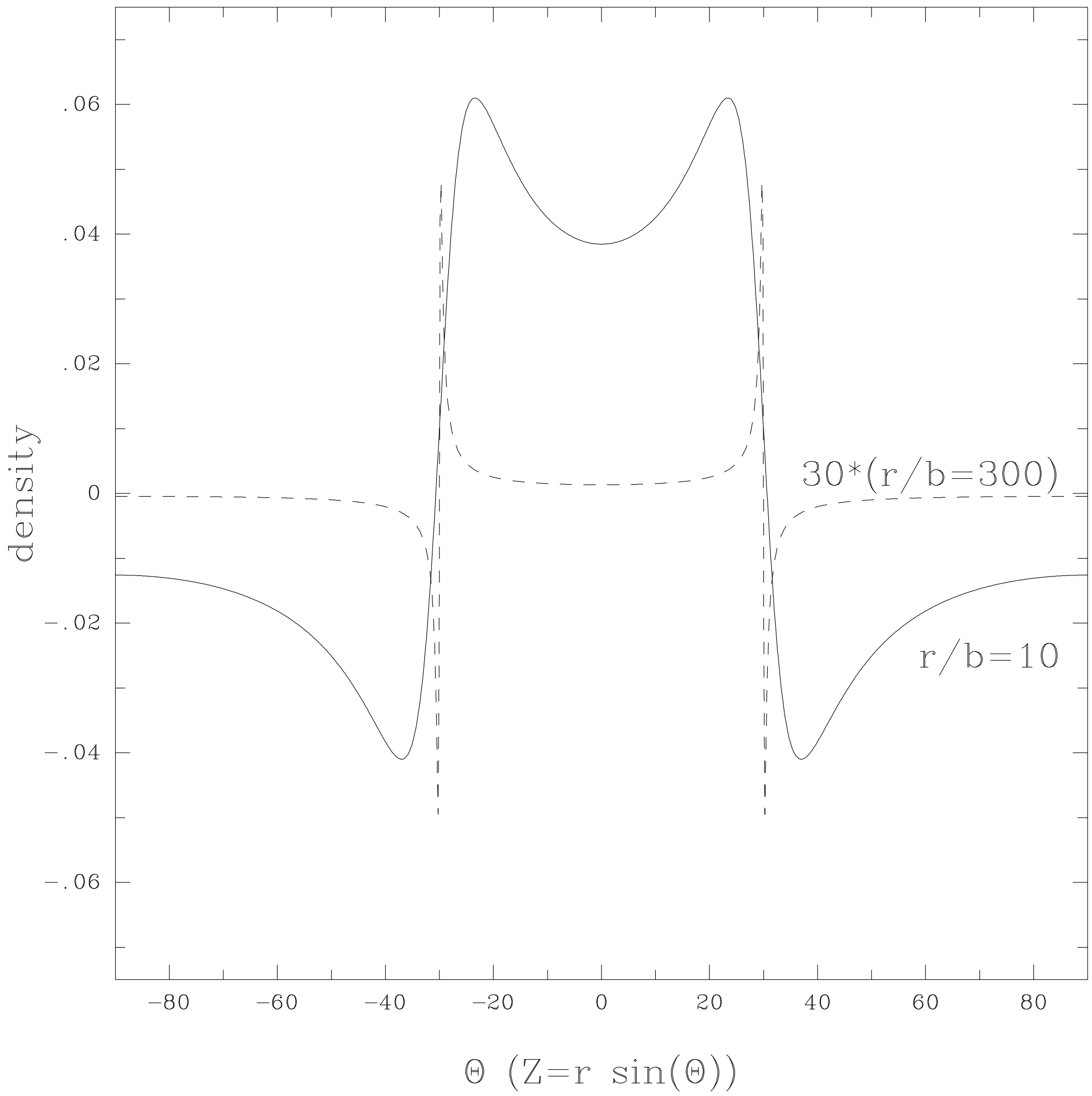

Fig. 2.— The konus function $\rho_{1+}(r, z|30°)$ as a function of angle ($z = r \sin\Theta$) for $r/b = 10$ (solid) and $r/b = 300$ (dashed) showing the development of the peaks straddling the curves $\alpha_{\pm} = 0$. The opening angle of the cone of uncertainty is 30°. The $r/b = 300$ curve was multiplied by the ratio of the radii for the two curves (30) to emphasize the different asymptotic rates.

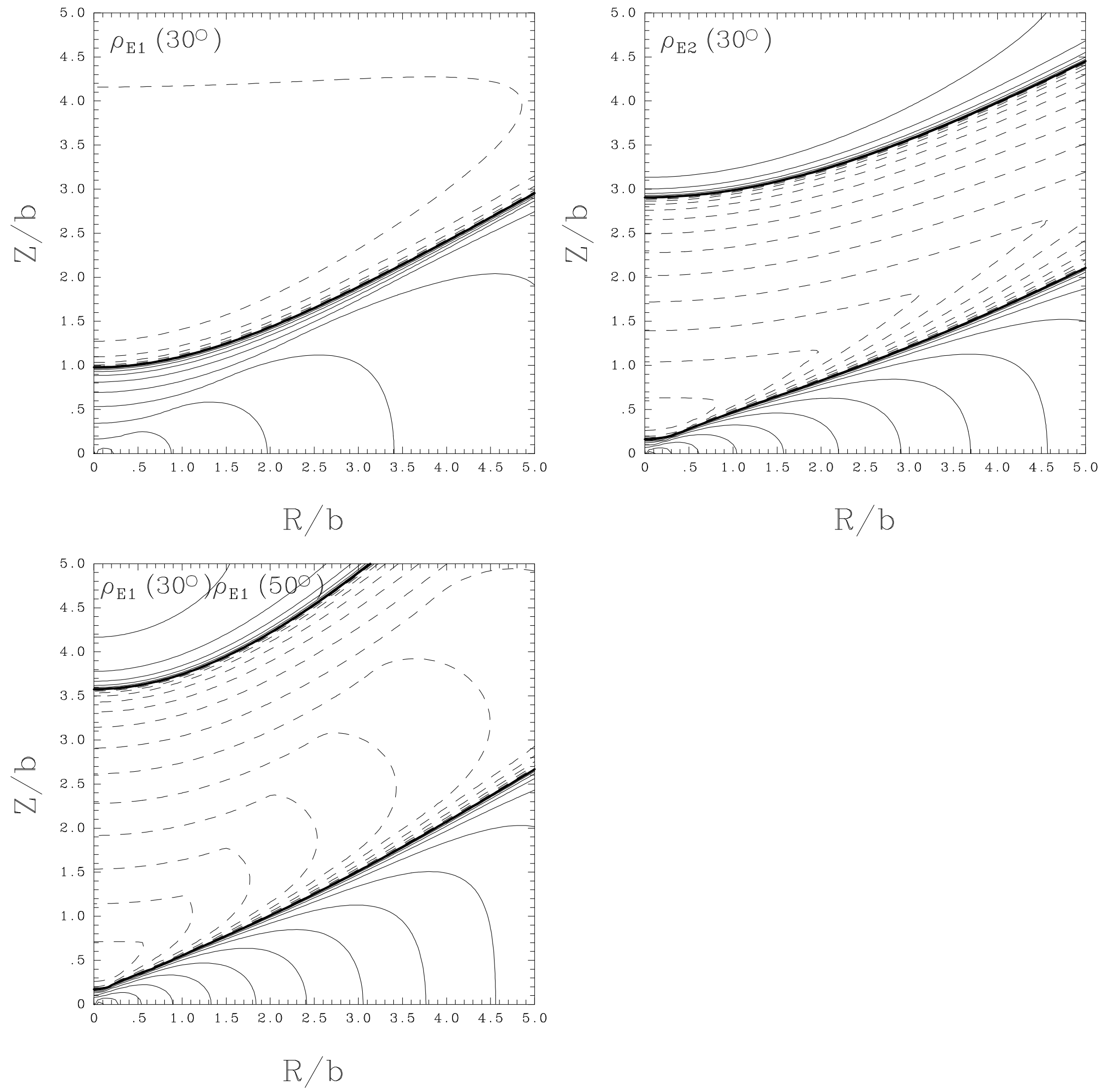

Fig. 4.— The top left, top right, and bottom right figures show contours of the exponential konus functions $\Re\rho_{E1}(r,z|30°)$, $\Re\rho_{E2}(r,z|30°) = \Re\rho^2_{E1}(r,z|30°)$, and $\Re\rho_{E1}(r,z|30°)\rho_{E1}(r,z|50°)$. The first two functions are asymptotically exponential except near $\alpha_\pm = 0$ where they decline as $r^{-1}$ and $r^{-2}$ respectively, while the last function is asymptotically exponential at all angles. The scaling and contours are the same as in Figure 1.